\documentclass[12pt]{iopart}
\usepackage{graphicx}
\usepackage{hyperref}
\usepackage{appendix}

\begin{document}

\title[Spectrally Resolved Higher Order Photon Statistics of SPDC]{Spectrally Resolved Higher Order Photon Statistics of Spontaneous Parametric Down Conversion}

\author{Jeffrey Carvalho$^*$, Chiran Wijesundara$^*$ \& Tim Thomay$^*$}

\address{$^*$Department of Physics, University at Buffalo - The State University of New York, Buffalo, NY 14260, USA}
\ead{thomay@buffalo.edu}
\vspace{10pt}
\begin{indented}
\item[]May 28, 2025

\end{indented}

\begin{abstract}

The photon statistics of Spontaneous Parametric Down Conversion (SPDC) exhibit dependencies on wavelength, pump power, and coincidence time. Notably, the average photon numbers were found to asymmetrically increase with increasing pump power around the degenerate wavelength of emission. By the coupling of the detection scheme to a spectrometer, studying different bandwidths within the emission revealed that shorter wavelengths increased nonlinearly with pump power, while longer wavelengths showed more linear behavior, indicating a wavelength dependent efficiency in the generation of the SPDC. We employ the use of a four detector Hanbury Brown and Twiss Interferometer to study the photon statistics of the signal beam, where the idler serves as the herald. The measured statistics were found to be best described by a Negative Binomial Distribution, which is a characteristic of thermal light sources. The detection and characterization of complex light sources has wide ranging applications in the fields of quantum metrology, quantum communications, and quantum computing, more specifically, a system that is sensitive to wavelength and photon number distribution. 

\end{abstract}

%
\vspace{2pc}

\section{Introduction}

Sources of higher order photon states have been of great interest in the fields of quantum communications \cite{sonoyama_generation_2024}, metrology \cite{deng_quantum-enhanced_2024}, and sensing \cite{tritschler_optical_2024}.
In communications, these states offer the ability to enhance security and information capacity \cite{cozzolino_high-dimensional_2019}. In metrology and sensing, quantum (Sub-Poissonian) states of light allows for operation below the shot-noise limit, as is well known, and this affect can be pushed further with higher-order photon states toward the Heisenberg limit \cite{deng_quantum-enhanced_2024}. However, the generation/manufacturing and detection of these states can be experimentally cumbersome \cite{dellanno_multiphoton_2006,xu_optimized_2024}. 

One common source for the generation of quantum states of light is Spontaneous Parametric Down Conversion (SPDC) \cite{euler_spectral_2021,harder_single-mode_2016,ma_highly_2023}. SPDC is a nonlinear process by which a pump photon interacting with a nonlinear crystal can spontaneously decay into two entangled photons (signal and idler) \cite{couteau_spontaneous_2018}. The SPDC photons posses a number of degrees of freedom in their generation, including frequency, polarization, time, and space \cite{ortega_spatial_2023}. While the polarization aspect is still widely studied \cite{kulkarni_intrinsic_2016,chaisson_phase-stable_2022,li_experimental_2023}, exploiting the spatial and temporal correlations between the down converted beams has shown to have the capability of providing higher degrees of entanglement than that of only using their polarization \cite{yorulmaz_role_2014}. There are many experimental factors that influence the properties of SPDC, such as the transverse pump beam modes \cite{walborn_spatial_2010,euler_spectral_2021}, pump focusing conditions \cite{suzer_does_2008,coccia_optimal_2023,lee_spatial_2016}, collection optics \cite{bennink_optimal_2010,bernecker_spatial_2023,sevilla-gutierrez_spectral_2024}, crystal parameters \cite{grice_spectral_1997,ramirez-alarcon_effects_2013}, and spatial filtering of the SPDC \cite{van_exter_effect_2006}. Since these correlations have been studied rigorously, the design parameters of the SPDC can be finely tuned, for example, maximizing the purity and/or generation rate of the down conversion \cite{rockovich_maximizing_2024,zhang_spontaneous_2021}. When the conversion efficiency is increased, there comes a turning point where multiple pairs of entangled photons are generated at the same time \cite{thekkadath_gain-induced_2024}. Depending on the application, the highest single pair generation rate may be of interest, or the multi photon conversions may be suppressed or enhanced \cite{lasota_optimal_2020,takeoka_full_2015}.

As previously demonstrated \cite{mauerer_how_2009}, there are different factors which can influence the photon statistics of the SPDC beams. For example, the spectral modes \cite{mauerer_how_2009}, spatial modes \cite{unternahrer_coincidence_2016}, temporal modes \cite{cohen_measuring_2023}, and pump statistics \cite{meher_dependence_2020} dependencies have been investigated for different configurations of SPDC. Depending application and designed experiment, the observed statistics can range from Sub-Poissonian \cite{waks_direct_2004}, which is typically exhibited by a quantum light source \cite{glauber_coherent_1963}, to Super-Poissonian \cite{paleari_thermal_2004}, which is typically exhibited by a thermal or chaotic source. Furthermore, depending on the modes present, multi mode down conversion can tend to a Poissonian, as the convolution of multiple thermal modes produces a Poissonian, or can lie in between a Poissonian and Super-Poissonian. \cite{avenhaus_photon_2008,mauerer_how_2009,cohen_measuring_2023} (See Section \ref{sec:distributions}).

While the photon statistics of SPDC light have been studied in many cases \cite{hockel_direct_2011,tapster_photon_1998,waks_direct_2004,waks_highly_2006,galinis_photon_2012,hamar_non-classical_2014,wasilewski_statistics_2008,arahata_wavelength_2021,dorfman_photon_2012}, more specifically, the joint photon counting statistics have been shown to described by a Poissonian Distribution \cite{schneeloch_introduction_2019,osadko_sub-_2007,trost_photon_2020,chan_role_2012}, while the signal (or idler) alone can exhibit a Thermal distribution if the measurement time scale is less than the coherence time \cite{kim_photon-counting_2022,kuhn_photon-statistics-based_2016,ou_multi-photon_2007,zmuidzinas_thermal_2003}. Since most times the detection time does not meet this requirement, spectral filtering can also cause the statistics to shift anywhere between these two extreme cases \cite{ou_multi-photon_2007,mauerer_how_2009,tapster_photon_1998}.

In the present work, we aim to generate the maximum number of entangled photon pairs via SPDC, and characterize the average photon number of the signal beam as functions of wavelength, pump power, and coincidence time, where the idler beam serves as the herald. With the capability of counting photon states up to $n=3$ without the use of number resolving detectors, this work serves to lay the ground work for characterizing more complex states, providing valuable information where the use of number-resolving detectors is not a major factor. 

Moreover, being able to characterize the photon number as functions of various parameters can prove useful in quantum sensing and quantum imaging, particularly in a system sensitive to wavelength and photon number distribution. Characterization of the photon number as functions of the studied parameters can also prove vital for the design and optimization of future complex non-classical light sources.


\section{Methods}

\begin{figure}
    \centering
    \includegraphics[width=0.75\linewidth]{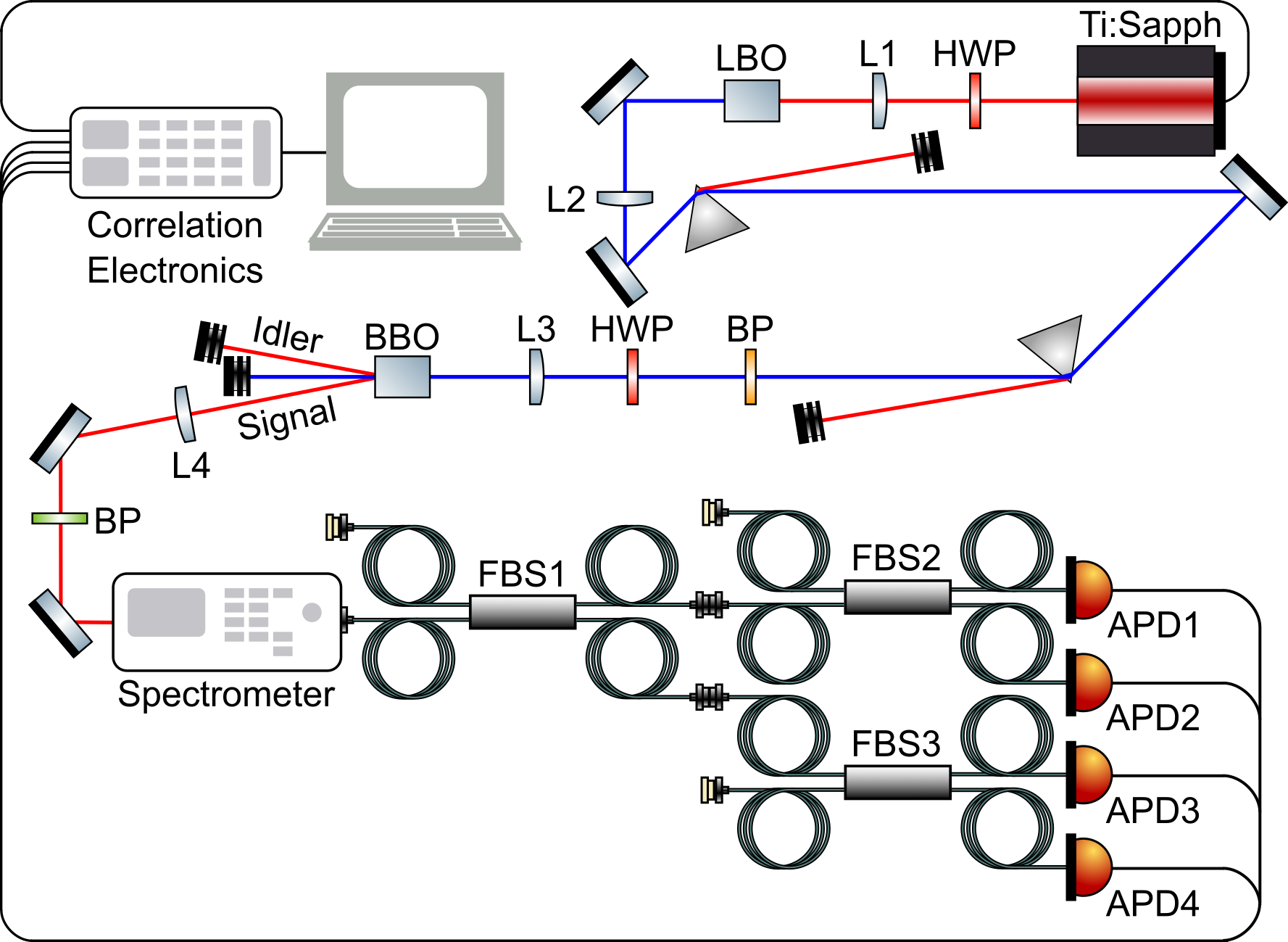}
    \caption{Experimental setup for measuring the photon statistics of SPDC coming from the $\beta$-BBO crystal. The time dynamics are investigated by synchronizing the APD events with the excitation laser in Time-Correlated Single Photon Counting (TCSPC) fashion via an FPGA based correlation board. The spectral dependence is studied by the use of the spectrometer before the HBT where the grating can be used to select what wavelength range coupled into the input fiber of the HBT. $APD$ Avalanche Photo Diode, $HWP$ half-wave plate, $BP$ band-pass filter, $FBS$ fiber beam splitter (50:50 ratio), $L$ lens.}
    \label{fig:setup}
\end{figure}

The experimental setup is shown in Figure \ref{fig:setup}. The SPDC is generated from a $1$mm thick $\beta$-BBO crystal, which is mounted on a rotational (translation) stage to optimize phase matching (focusing). Since the crystal is type-I ($e\rightarrow o+o$), the signal and idler photons will have the same polarization, that is, perpendicular to the pump polarization. The signal and idler are spatially separated by means of non-collinear phase matching, so that they form a cone around the pump beam.  

The pump ($400$ nm) beam used is generated by frequency doubling (Second Harmonic Generation) the output of a pulsed (250 kHz, 270 fs) Ti:Sapph laser. The SHG is focused into the $\beta$-BBO crystal with an $f_3=100$ mm ($W_p=146\pm3$ $\mu$m FWHM) lens. Here we are interested in studying around the degenerate emission, meaning $\lambda_s = \lambda_i = 800$ nm, so a band pass filter (40 nm FWHM) is used. The pump beam is also spatially blocked as in Figure \ref{fig:setup}. The signal beam is then collected by an $f_4=100$ mm lens ($W_s=189\pm10$ $\mu$m FWHM) and directed into a spectrometer in order to study the spectral dependence of the photon statistics.  

To measure photon statistics, a four detector (Avalanche Photo Diodes) graded index (GRIN) multi mode fiber-based Hanbury Brown and Twiss Interferometer \cite{brown_correlation_1956} is used, enabling the counting of photon events up to $n=3$. Since there are possibilities of all four detectors recording an event at the same time, there are recorded $n=4$ events which are treated as $n=4+$ as the detectors are not number-resolving. The measurements are taken in a Time-Correlated Single Photon Counting (TCSPC) fashion, so LVTTL signals from the APDs and excitation laser (called reference pulses) are recorded and time-tagged by an FPGA correlation board. The relative time difference between events ($\Delta\tau$) is used to count coincidence detections within the same reference pulse. In post-collection, $\Delta\tau$ is varied to investigate its effect on the photon statistics, allowing for the probing of different temporal modes present in the detected photon events. The range of $\Delta\tau$ used is from 0.165 ns (minimum resolution of the correlation board) up to 0.823 ns. 

To investigate the spectral dependence of the statistics, the input lead of the fiber HBT is Numerical Aperture (NA) matched to the output of a spectrometer. The spectrometer used contains 50 lines/mm grating, the central wavelength of which is varied in order to change the portion of the SPDC spectrum that is coupled into the HBT setup. The entrance slit width is also set to 50 $\mu$m for the measurements. The ranges of wavelengths ($\lambda$) studied goes from 783 nm to 819 nm in steps of 4 nm, so a total of ten wavelengths are measured for each pump power. The average pump power is varied by use of ND filters, and the measured average powers used were approximately 39 mW (156 nJ), 33 mW (132 nJ), 25 mW (100 nJ), and 14 mW (56 nJ). The SPDC signal started to be on the same order as the noise level at the tested powers lower than 14 mW.

Using the approach in \cite{rockovich_maximizing_2024}, the signal rate for the experimental parameters in this work was calculated and used in conjunction with the measured events to derive the single photon detection efficiency of the setup, which was found to be $\eta_{setup}=12\pm2\%$. A large factor of uncertainty comes from the quantum efficiency of the APDs, specified by the manufacturer as $7\%$ at 800 nm. Over the wavelength range studied from 780 nm to 820 nm, this value fluctuates by $\pm1\%$, and when propagated contributes significantly to the $\pm2\%$ of the overall system detection efficiency. $\eta_{setup}$ is used to scale the detected photon events to account for the detection losses present in the setup. The number of $n=0$ events are derived from the difference between the total detection opportunities in the measurement and the scaled total detection events, and the overall normalization for each dataset is the total detection opportunities. The $n=1,2,3$ photon events are corrected for the non number resolving capability of the APDs by using the quantum mechanical probabilities of higher order states triggering events of lower orders at the subsequent number of detectors, as detailed in Section \ref{sec:correction}. 

\section{Results}

The measured spectrum of the SPDC was found to be asymmetric about the degenerate wavelength of 800 nm, with lower wavelengths having higher intensity. Asymmetric spectra have also been shown in previous works (for example \cite{baek_spectral_2008}) and is sometimes attributed to loss in the setup \cite{spasibko_spectral_2020}, for example, detection inefficiency or optical transmission. The asymmetry could also arise from group velocity mismatch, where the velocity differences between the pump and SPDC photons can distort the phase matching conditions leading to different conversion efficiencies across the SPDC spectrum \cite{ou_multi-photon_2007}. As the pump power decreased, the asymmetry appeared to be less prominent, which can be explained as the overall conversion efficiency is decreased, the group velocity dispersion effect is less pronounced.  

\begin{figure}
    \centering
    \includegraphics[width=1.0\linewidth]{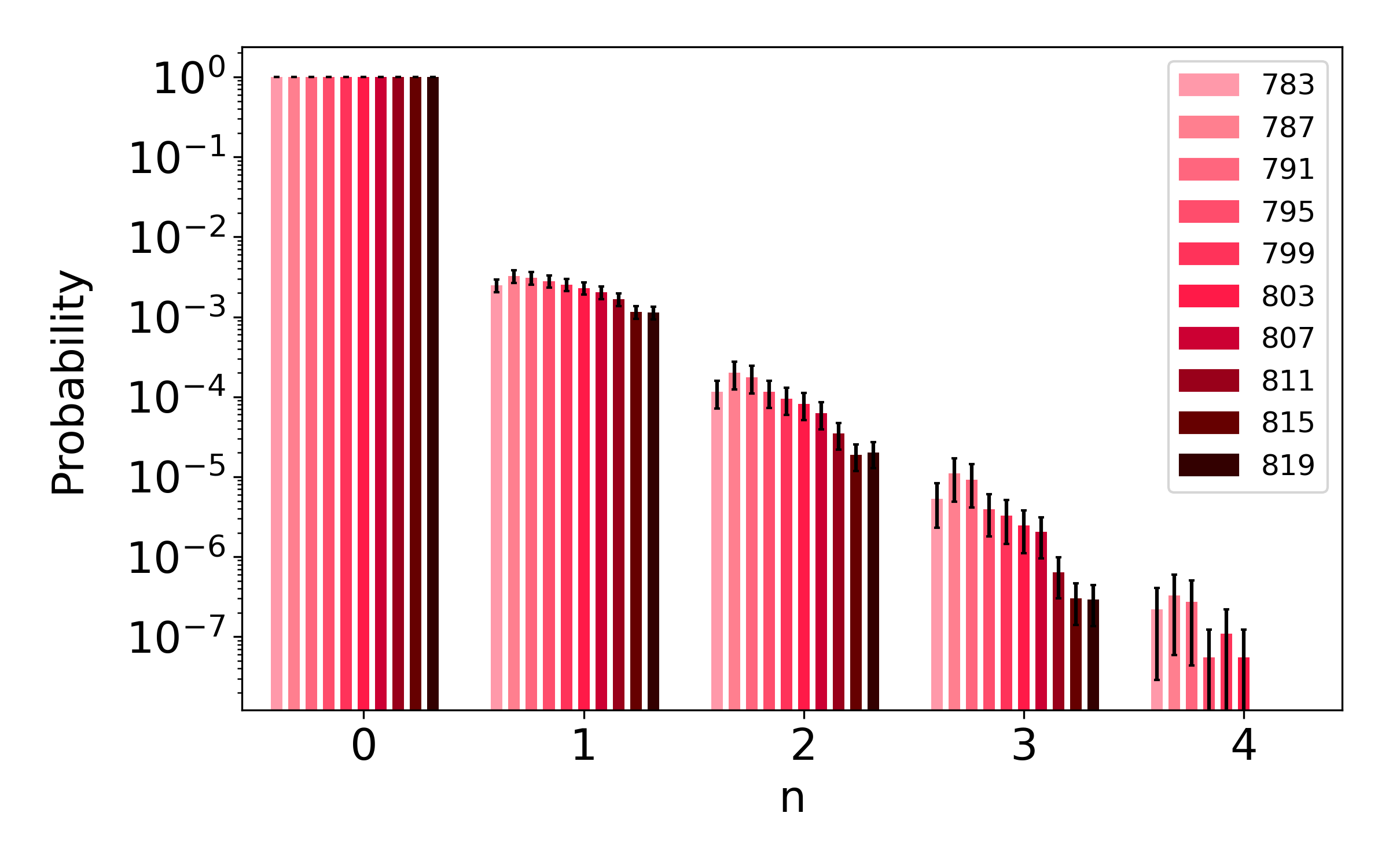}
    \caption{Spectrally resolved photon statistics for the ten measurements using the highest average pump power ($\approx39$ mW) and lowest coincidence window (0.165 ns). The x axis represents photon number $n$ and the y axis represents probability, while the colors of the bars indicate the central wavelength of the grating in the spectrometer before the HBT setup in Figure \ref{fig:setup}. The intensity profiles of the distributions seem to closely follow that of the observed spectrum, where the wavelengths lower than 800 nm show higher probabilities of photon states $n>1$}
    \label{fig:fullpsp40t0}
\end{figure}

Figure \ref{fig:fullpsp40t0} shows the spectrally resolved photon probability distributions for the 10 studied wavelengths using the lowest coincidence window $\Delta\tau=0.165$ ns and highest average pump power $\bar{P}\approx39$ mW. The spectral dependence of the distributions follows closely to that of the asymmetric spectrum, where the highest probability of $n>1$ events occurs at the highest intensity wavelength of 787 nm. These individual probabilities are fit with a negative binomial as explained in the previous section to extract their respective average photon numbers.

\begin{figure}
    \centering
    \includegraphics[width=1.0\linewidth]{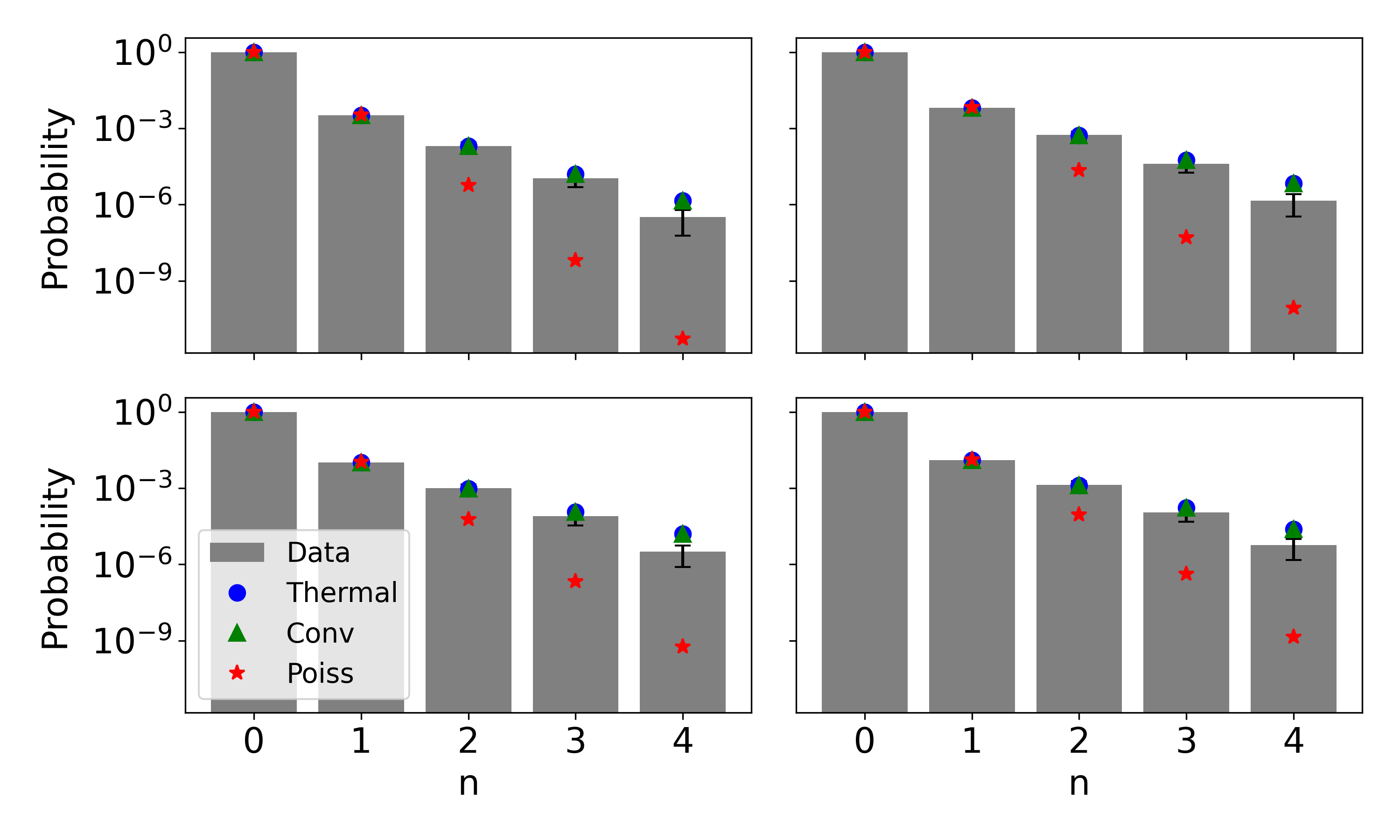}
    \caption{Fitting results for the central grating wavelength of 787 nm with the highest average pump power. On the x axis is photon number detection $n$ and probability is on the y axis. The four plots represent four of the coincidence windows used. From left to right the coincidence values are 0.165, 0.329, 0.494, and 0.658 ns. The gray bars represent the normalized experimental data, the red stars is the Poissonian fit, blue circles is the Negative Binomial fit (thermal), and the green triangles is the convolution of the two.}
    \label{fig:fitsplot}
\end{figure}

To quantitatively compare the variations in photon statistics as functions of wavelength, power, and time, the probability distributions are fit to extract the average photon number $\langle n \rangle$. The choice of fitting function inherently reveals information about the modal complexity of the SPDC. Depending on the modal complexity, the statistics can be best described by a Poisson Distribution, Negative Binomial (Thermal) Distribution, or a convolution of the two \cite{avenhaus_photon_2008,mauerer_how_2009,cohen_measuring_2023}. The statistics are thus fit with all three of these distributions to see which of them best describes the observed SPDC. The distributions are fit out to $n=8$ in the effort to capture the tail of the distribution.

Figure \ref{fig:fitsplot} shows the fitted distributions for the central grating wavelength of 787 nm at the highest average pump power for four of the coincidence time windows used, where the largest coincidence time window is excluded for simplicity. Comparing the the fits visually, the Negative Binomial is seems to be the best fit to the data, where the Poisson underestimates the higher order states $n>1$. Comparing the $R^2$ scores when $\Delta\tau=0.165$ ns, due to the $n=0$ dominance of the data, the values are very close to 1 for all of the fits, but comparing the thermal $R^2$ to the Poissonian one, the thermal $R^2$ is larger by approximately 7E-08, whereas the difference between the thermal and convolution was 3E-16. Furthermore, when $\Delta\tau=0.658$ ns, the difference between the thermal $R^2$ and Poissonian and the thermal and convolution was 3E-06 and 7E-12, respectively.

The Poissonian Distribution describes a more random process, while the Negative Binomial corresponds to correlated events. Since the Negative Binomial seems to best describe the data, the photon arrival times seem to be strongly correlated. The improved fit of the Negative Binomial, particularly at higher order states, also speaks to the spectral filtering by the spectrometer before coupling into the HBT setup. The effective bandwidth of selection that couples into the fiber is narrow enough to show the bunching effect of the thermal distribution \cite{mauerer_how_2009,tapster_photon_1998}.

An interesting piece of information is seen through comparing the goodness of fits while increasing the coincidence window. Comparing the first plot to the fourth, the Negative Binomial is still the best fit, but the Poissonian is visually a much better fit in the maximum time window shown than in  the minimum case. However, comparing the Poisson $R^2$ in the two cases, the difference between the $\Delta\tau=0.165$ ps and $\Delta\tau=0.658$ ns datasets is 3E-06, indicating that based on the $R^2$ the Poissonian fits better with a shorter coincidence window. This is likely due to the zero dominance of the data, where when $\Delta\tau=0.658$ ns the fit to the zeros may be worse than in the $\Delta\tau=0.165$ ns case, but the fits to the higher order states is better, as shown in Figure \ref{fig:fitsplot}. 

Since all of these photons are being generated from within the same laser pulse, and the laser pulse time scale is so much shorter than the achievable detection resolution, this behavior suggests that even though the events within the pulse are correlated, the strength of correlation seems to decrease as events separated farther apart in time are correlated. This decrease in correlation strength could be due to the temporal modes effectively averaging each other out, which has been shown previously when correlating pairs far apart from one another in time. This increase in coincidence window effectively causes a loss of ability to distinguish between the finer temporal structure of pulses arriving at the detectors \cite{blauensteiner_photon_2009}. This can also be due to some experimental parameters, and since the time resolution is not fine enough to resolve correlations at the time scales of the laser pulse, probing the temporal modes within the detection is interesting. Since the Negative Binomial best describes the data, it is used to extract the average photon number, as a function of its input parameters $r$ the number of failures and $p$ the probability of success. The average photon number is then calculated as $\langle n\rangle=r(1-p)/p$. 

\begin{figure}
    \centering
    \includegraphics[width=1.0\linewidth]{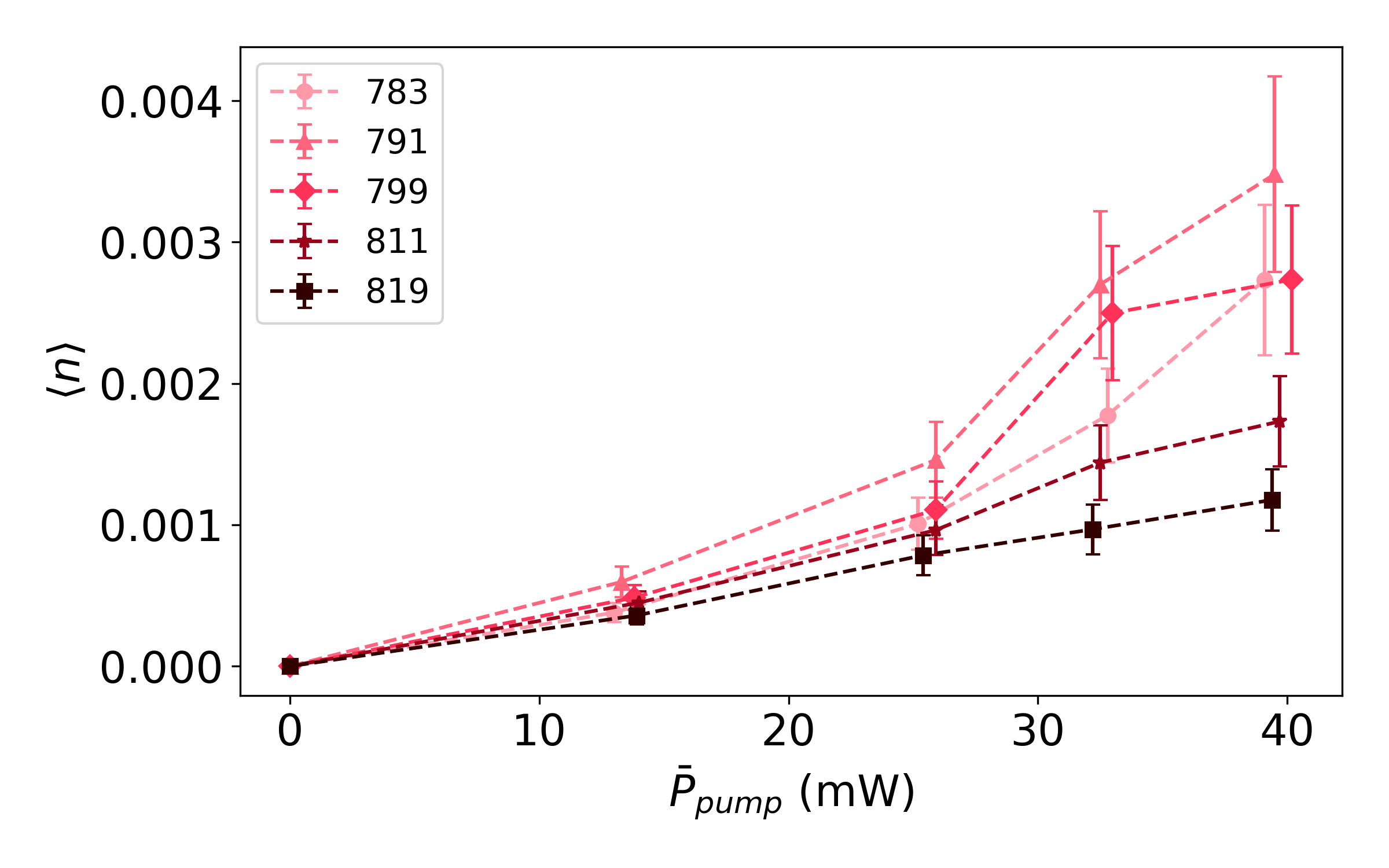}
    \caption{Average photon number power dependence for five of the ten measured wavelengths. The x axis represents the pump power and the y axis represents the average photon number extracted from the Negative Binomial fit. Spectral dependence is also shown by the color in the legend. A nonlinear relationship is shown at the shorter wavelengths, where the conversion efficiency is higher. The maximum intensity wavelength found was 787 nm, which shows the most nonlinear relationship. For wavelengths more than 800 nm, a more linear relationship is observed.}
    \label{fig:avgnpower}
\end{figure}

When examining the average photon numbers, $\langle n\rangle_\lambda$ refers to the average photon number for grating wavelength $\lambda$. Figure \ref{fig:avgnpower} shows the average photon number as a function of pump power for five out of the ten measured wavelengths for the lowest coincidence time of 0.165 ns, where the zero serves as a baseline. Since the phase matching of the crystal is optimized to the highest counts, due to the asymmetry observed in the spectrum, the optimized wavelength of 787 nm shows the highest average photon number at the highest average pump power of $\approx39$ mW. In this region, there is a clear nonlinear relationship between the photon number and pump power, indicating more efficient generation of SPDC photons. This nonlinear relationship comes from the fact that at these wavelengths, since the efficiency is higher, the probability of generating photon states with numbers $n>1$ increased, causing the average photon number to also increase nonlinearly. In comparison, there is more linear behavior where $\lambda=819$ nm, as it is farther away from the optimized phase matching peak, and thus its conversion efficiency is lower.  In comparison of wavelengths, at the highest average pump powers, $\langle n\rangle_{787}=0.0037\pm 0.0007$ and $\langle n\rangle_{819} =0.0012 \pm 0.0002$, showing an approximate 3.1-fold decrease, which is also consistent in the intensity variations.

\begin{figure}
    \centering
    \includegraphics[width=1.0\linewidth]{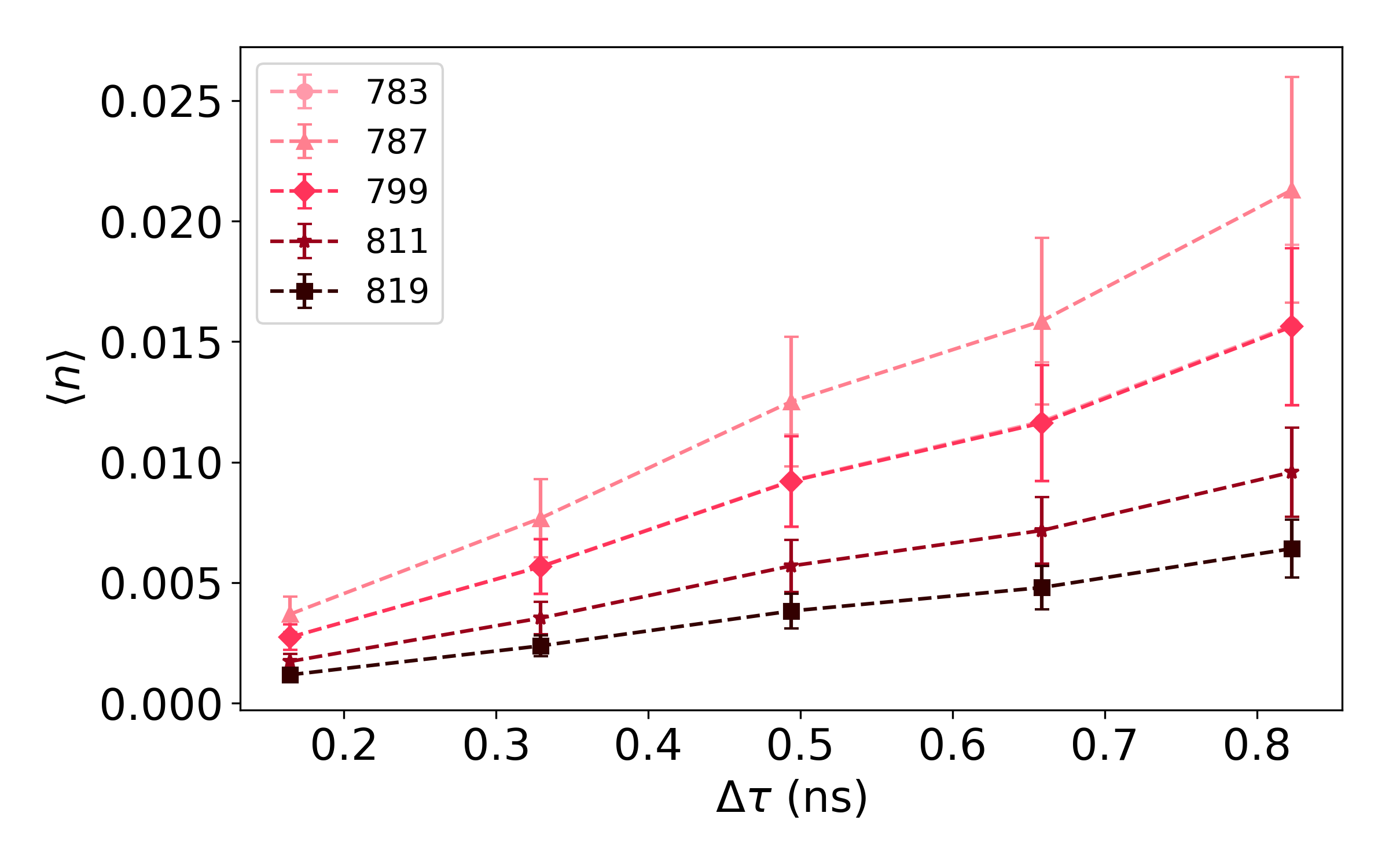}
    \caption{Time dependence of the average photon number $\langle n\rangle$ for five out of the ten measured wavelengths. The x axis represents the coincidence window and the y axis represents the average photon number. The time dependence is studied by varying the coincidence window $\Delta\tau$ to correlate photon arrival times. A steady relative increase is shown for $\Delta\tau<0.494$ ns, and a saturation effect is shown between 0.494 ns and 0.658 ns. The steeper increase from 0.658 ns to 0.823 ns can be explained by the asymmetric response function of the APDs used in the measurement, where the FWHM of the response function was found to be approximately 0.658 ns.}
    \label{fig:avgntimep40}
\end{figure}

At the lowest pump power of $\approx14$ mW, $\langle n\rangle_{787} = 0.00054\pm 0.00010$, showing an increase of approximately 6.9-fold when the pump power increases from $\approx14$ mW to $\approx39$ mW. In comparison, $\langle n\rangle_{819} = 0.00036\pm 0.00006$ with $\approx14$ mW pump power, showing an increase of approximately 3.3-fold in the average photon number. These differences show the rapidly increasing generation of the SPDC at $\lambda=787$ nm, and show the more linear relationship between the pump power and average photon number at $\lambda=819$ nm. The observed flattening of $\langle n\rangle _{799}$ at the highest power can most likely be attributed to experimental error, possibly related to laser stability. While this type of laser is usually short term stable, over long periods of time, like the time periods of these measurements, the laser can become unstable in aspects such as mode hopping. In the interest of being able to reliably compare the data across the data sets, the counts were kept at a maximum and restored when an observed decrease in counts was shown. When increasing the coincidence time, the absolute average photon numbers changed but the overall trends remained the same in this case.

Figure \ref{fig:avgntimep40} shows the dependence of $\langle n\rangle$ on the relative time difference window between detection events $\Delta\tau$ for the highest average pump power. As shown in Figure \ref{fig:fitsplot}, in increasing the coincidence window, the number of detection opportunities decrease and the number of higher order states increase, effectively shifting the photon number distribution to the right, i.e. higher photon number. Since the measurements are taken in TCSPC fashion, and the down converted photons are produced by the same laser pulse, they are strongly correlated, as exhibited by the Negative Binomial fit best describing the data. The relationship between average photon number and coincidence window is mostly linear, as shown in Figure \ref{fig:avgntimep40}.  As the coincidence window approaches 0.658 ns, a slight saturation effect in the average photon is observed when compared to the shorter time windows, suggesting that increasing the window from 0.494 to 0.658 ns adds less correlated events than the smaller windows chosen. Following the saturation effect is a slight increase in slope from changing $\Delta\tau$ from 0.658 ns to 0.823 ns. This is likely from the asymmetric instrument response function from the combination of the APDs and correlation board used in this measurement. The response function was found to have a FWHM of approximately 0.823 ns, with an asymmetry in the slow decaying tail versus the relatively faster rising edge. As the coincidence is increased to the FWHM, the asymmetric response function seems to start to correlate events from the rising to the falling edge, causing the sharper increase in all of the cases shown in Figure \ref{fig:avgntimep40}.

However, the difference in overall slope based on wavelength is noticeable, where the shorter wavelengths $\lambda<800$ nm show more rapid increase in average photon number as the $\Delta\tau$ increases. This can be explained by the fact that at the lower wavelengths, as shown previously, the efficiency of generation is higher, leading to more detected events overall. When increasing the coincidence window, there is a more pronounced increase due to the higher number of generated photons. This result shows in comparing the longer time windows for correlation between the arrival times of the generated SPDC photons, the average photon number increases regardless of wavelength. Correlating events far beyond the 0.823 ns window chosen here does not add much additional information, and would be adding increasingly more dark counts (noise) since the events are relatively localized in time. 

\section{Conclusion}

This work serves to lay the ground work of characterizing and optimizing more complex light sources of higher order photon states. The presented results examined the photon statistics dependence of type-I degenerate noncollinear Spontaneous Parametric Down Conversion as functions of wavelength, pump power, and coincidence time. The detection scheme used employs a four detector Hanbury Brown and Twiss Interferometer \cite{brown_correlation_1956} using multi mode fiber with the capability of being used across the visible spectrum.

The SPDC emission was found to have asymmetric generation efficiency at high pump powers around the degenerate wavelength of 800 nm, where the lower wavelengths proved to have a higher intensity. This behavior was transmitted directly to the subsequent photon statistics, where the highest probability of coincidence detection events was observed at 787 nm, while the lowest probability was around 819 nm. The photon events were found to be strongly correlated, as their probability distributions were found to be best described by a Negative Binomial distribution, which commonly characterizes a thermal light source. 

The average photon numbers were extracted from the fittings of the distributions to quantitatively compare their dependence on the studied parameters. In wavelength, the dependence was found to follow the asymmetric behavior of the intensity at high pump powers, with 787 nm showing the highest average photon number and 819 nm showing the lowest. As the pump power was decreased, the asymmetric behavior flattened out, showing a more symmetric distribution across the spectrum. In power, the average photon number was found to increase nonlinearly with increasing pump power at wavelengths around 787 nm, while the dependence was more linear around 819 nm, which can be explained through the higher efficiency of generation at these wavelengths. In time, the behavior was mostly linear, with a slight saturation effect shown using a coincidence window of 0.658 ns. Beyond this window a sharp linear increase was found, likely owing to the asymmetric response function of the detection setup.

These findings contribute valuable insights into the characterization of the SPDC photon statistics and hold promise in the developments of quantum sensing and imaging systems, particularly in a system sensitive to wavelength and photon number distribution. 

\section{Acknowledgments}

The authors are greatly appreciative of the assistance and computational implementation provided by Andrew Rockovich \cite{rockovich_maximizing_2024} at the end of the calculation of the theoretical photon emission rate used to extract the detection efficiency and normalize the data. 

\section{Conflict of Interest Statement}

The authors claim no conflict of interest.

\section{Data Availability Statement}

The experimental data used to make the figures will be publicly accessible through Dryad when the link becomes available.

\section{Supplemental}\label{sec:supplemental}
\subsection{Photon Probability Distributions}\label{sec:distributions}

Because of the bosonic behavior of the SPDC emission, correlations within one of the beams can be bunched, leading to thermal statistics \cite{blauensteiner_photon_2009,hockel_direct_2011}. In the case of thermal light, the statistics are governed by the Bose-Einstein Distribution \cite{fox_quantum_2006}:

\begin{equation}\label{eq:bedist}
    P(n)=\frac{1}{\langle{n}\rangle+1}\left(\frac{\langle{n}\rangle}{\langle{n}\rangle+1}\right)^n 
\end{equation}

\noindent where $\langle n\rangle$ is the average photon number, which exhibits the characteristic behavior of a variance greater than the mean, $(\Delta n)^2=\langle n\rangle+\langle n\rangle^2$ \cite{wasilewski_statistics_2008}. In practice, the bunching effects of the down conversion can be masked by various experimental parameters, for example, detection resolution if the resolving times are much longer than the coherence time of the down conversion \cite{blauensteiner_photon_2009}. Additionally, since the down conversion can be emitted into multiple modes, the correlations between the modes \cite{avenhaus_photon_2008} can also cause the statistics to tend toward Poissonian, which is \cite{fox_quantum_2006}: 

\begin{equation}
    P(n)=\frac{\langle {n}\rangle ^n}{n!}e^{-\langle{n}\rangle} 
\end{equation}

\noindent which has the characteristic behavior of a variance equal to the mean, $(\Delta n)^2=\langle n\rangle$. Conceptually, the thermal statistics correspond to events that are highly correlated, while the Poisson describes a random process. 

As the modal complexity of the down conversion changes, like having more spectral or spatial modes, the statistics can fluctuate between thermal  \cite{paleari_thermal_2004}, Poisson \cite{avenhaus_photon_2008}, and a convolution of the two \cite{mauerer_how_2009,cohen_measuring_2023}. In practice, spectral filtering can be employed to ensure the collection of a single mode \cite{ou_multi-photon_2007,tapster_photon_1998}. 

\subsection{Non-number Resolving Correction}\label{sec:correction}

Since the APDs are inherently non photon number resolving detectors, recording a single event technically means there was at least one photon. In order to correct for the photon detection numbers, the quantum mechanical probabilities that a certain photon number higher than 1 can trigger a single event are calculated up to $n=4$. This was also done similarly for $n=2$ and $n=3$, so that the extracted probabilities reflect $n=1,2,3$ instead of $n=1+,2+,3+$. In order to find the probability that a certain input photon state can get through the detection setup, simulating the HBT of three 50:50 ratio beam splitters was carried out in Python, where the exits ultimately end up at the four detectors. The factors extracted are then multiplied by the corresponding detection probability for that photon number in order to correct for the probability that photon states higher than the detected photon number are present. The resulting probabilities are then:

\begin{equation}\label{eq:trueprobs}
    P(n)_{true}=\cases{
    P(1)_{meas}-\left[A_2 P(2)_{meas}+A_3 P(3)_{meas}+A_4P(4)_{meas} \right]& n=1\\
    P(2)_{meas}-\left[B_3 P(3)_{meas}+B_4 P(4)_{meas}\right]& n=2\\
    P(3)_{meas}-\left[C_4 P(4)_{meas}\right]& n=3\\
    P(4)_{meas}& n=4\\
    1-\sum_{n=1}^4 P(n)_{true}& n=0
    }
\end{equation}

\noindent where the factors $A_2,A_3,A_4$ correspond to input states of $|2\rangle,|3\rangle,|4\rangle$ making it through the beam splitter network and triggering a single photon event, $B_3,B_4$ correspond to input states of $|3\rangle,|4\rangle$ making it through the network and triggering a two photon event, and $C_4$ corresponds to a $|4\rangle$ input state making it through the network and triggering a three photon event. For example, for input states of $|2\rangle,|3\rangle,|4\rangle$, triggering a single photon event at any of the detectors corresponds to $A_2=0.25,A_3=0.0625,A_4\approx0.0156$, respectively. In the interest of using the measured data for the correction, the calculated input states are up to $|4\rangle$, and for input states higher the correction would be increasingly smaller. The associated errors in the counts and efficiency and thus the probabilities were propagated using the covariance matrix for Equation \ref{eq:trueprobs}. 
\section{References}
\bibliographystyle{unsrt}
\bibliography{manuscript}

\begin{thebibliography}{10}

\bibitem{sonoyama_generation_2024}
Tatsuki Sonoyama, Kazuma Takahashi, Tomoki Sano, Takumi Suzuki, Takefumi Nomura, Masahiro Yabuno, Shigehito Miki, Hirotaka Terai, Kan Takase, Warit Asavanant, Mamoru Endo, and Akira Furusawa.
\newblock Generation of multi-photon {Fock} states at telecommunication wavelength using picosecond pulsed light, May 2024.
\newblock arXiv:2405.06567 [quant-ph].

\bibitem{deng_quantum-enhanced_2024}
Xiaowei Deng, Sai Li, Zi-Jie Chen, Zhongchu Ni, Yanyan Cai, Jiasheng Mai, Libo Zhang, Pan Zheng, Haifeng Yu, Chang-Ling Zou, Song Liu, Fei Yan, Yuan Xu, and Dapeng Yu.
\newblock Quantum-enhanced metrology with large {Fock} states.
\newblock {\em Nature Physics}, 20(12):1874--1880, December 2024.
\newblock Publisher: Nature Publishing Group.

\bibitem{tritschler_optical_2024}
Patrick Tritschler, Torsten Ohms, André Zimmermann, Fabian Zschocke, Thomas Strohm, and Peter Degenfeld-Schonburg.
\newblock Optical interferometer using two-mode squeezed light for enhanced chip-integrated quantum metrology.
\newblock {\em Physical Review A}, 110(1):012621, July 2024.

\bibitem{cozzolino_high-dimensional_2019}
Daniele Cozzolino, Beatrice Da~Lio, Davide Bacco, and Leif~Katsuo Oxenløwe.
\newblock High-{Dimensional} {Quantum} {Communication}: {Benefits}, {Progress}, and {Future} {Challenges}.
\newblock {\em Advanced Quantum Technologies}, 2(12):1900038, 2019.
\newblock \_eprint: https://onlinelibrary.wiley.com/doi/pdf/10.1002/qute.201900038.

\bibitem{dellanno_multiphoton_2006}
Fabio Dell’Anno, Silvio De~Siena, and Fabrizio Illuminati.
\newblock Multiphoton quantum optics and quantum state engineering.
\newblock {\em Physics Reports}, 428(2):53--168, May 2006.

\bibitem{xu_optimized_2024}
Guangpeng Xu, Jeffrey Carvalho, Chiran Wijesundara, and Tim Thomay.
\newblock Optimized higher-order photon state classification by machine learning.
\newblock {\em APL Quantum}, 1(3):036122, September 2024.

\bibitem{euler_spectral_2021}
Sabine Euler, Erik Fitzke, Oleg Nikiforov, Daniel Hofmann, Till Dolejsky, and Thomas Walther.
\newblock Spectral characterization of {SPDC}-based single-photon sources for quantum key distribution.
\newblock {\em The European Physical Journal Special Topics}, 230(4):1073--1080, June 2021.

\bibitem{harder_single-mode_2016}
Georg Harder, Tim~J. Bartley, Adriana~E. Lita, Sae~Woo Nam, Thomas Gerrits, and Christine Silberhorn.
\newblock Single-{Mode} {Parametric}-{Down}-{Conversion} {States} with 50 {Photons} as a {Source} for {Mesoscopic} {Quantum} {Optics}.
\newblock {\em Physical Review Letters}, 116(14):143601, April 2016.

\bibitem{ma_highly_2023}
Zhaohui Ma, Jia-Yang Chen, Malvika Garikapati, Zhan Li, Chao Tang, Yong~Meng Sua, and Yu-Ping Huang.
\newblock Highly efficient and pure few-photon source on chip.
\newblock {\em Physical Review Applied}, 20(4):044033, October 2023.

\bibitem{couteau_spontaneous_2018}
Christophe Couteau.
\newblock Spontaneous parametric down-conversion.
\newblock {\em Contemporary Physics}, 59(3):291--304, July 2018.
\newblock arXiv:1809.00127 [physics, physics:quant-ph].

\bibitem{ortega_spatial_2023}
Evelyn~A. Ortega, Jorge Fuenzalida, Mirela Selimovic, Krishna Dovzhik, Lukas Achatz, Sören Wengerowsky, Rodrigo~F. Shiozaki, Sebastian~Philipp Neumann, Martin Bohmann, and Rupert Ursin.
\newblock Spatial and spectral characterization of photon pairs at telecommunication wavelengths from type-0 spontaneous parametric downconversion.
\newblock {\em Journal of the Optical Society of America B}, 40(1):165, January 2023.

\bibitem{kulkarni_intrinsic_2016}
Girish Kulkarni, V.~Subrahmanyam, and Anand~K. Jha.
\newblock Intrinsic upper bound on two-qubit polarization entanglement predetermined by pump polarization correlations in parametric down-conversion.
\newblock {\em Physical Review A}, 93(6):063842, June 2016.
\newblock Publisher: American Physical Society.

\bibitem{chaisson_phase-stable_2022}
Zachary M.~E. Chaisson, Patrick~F. Poitras, Micaël Richard, Yannick Castonguay-Page, Paul-Henry Glinel, Véronique Landry, and Deny~R. Hamel.
\newblock Phase-stable source of high-quality three-photon polarization entanglement by cascaded down-conversion.
\newblock {\em Physical Review A}, 105(6):063705, June 2022.
\newblock Publisher: American Physical Society.

\bibitem{li_experimental_2023}
Cheng Li, Boris Braverman, Girish Kulkarni, and Robert~W. Boyd.
\newblock Experimental generation of polarization entanglement from spontaneous parametric down-conversion pumped by spatiotemporally highly incoherent light.
\newblock {\em Physical Review A}, 107(4):L041701, April 2023.
\newblock Publisher: American Physical Society.

\bibitem{yorulmaz_role_2014}
S.~C. Yorulmaz, M.~P. van Exter, and M.~J.~A. de~Dood.
\newblock The role of spatial and temporal modes in pulsed parametric down-conversion.
\newblock {\em Optics express}, 22(5):5913--5926, 2014.
\newblock Place: WASHINGTON Publisher: Optical Soc Amer.

\bibitem{walborn_spatial_2010}
S.P. Walborn, C.H. Monken, S.~Pádua, and P.H. Souto~Ribeiro.
\newblock Spatial correlations in parametric down-conversion.
\newblock {\em Physics Reports}, 495(4-5):87--139, October 2010.

\bibitem{suzer_does_2008}
Özgün Süzer and Theodore G.~Goodson Iii.
\newblock Does pump beam intensity affect the efficiency of spontaneous parametric down conversion?
\newblock {\em Optics Express}, 16(25):20166--20175, December 2008.
\newblock Publisher: Optica Publishing Group.

\bibitem{coccia_optimal_2023}
Lorenzo Coccia, Alberto Santamato, Giuseppe Vallone, and Paolo Villoresi.
\newblock Optimal focusing conditions for bright spontaneous parametric down-conversion sources.
\newblock {\em Physical Review A}, 107(6):063712, June 2023.

\bibitem{lee_spatial_2016}
Jong-Chan Lee and Yoon-Ho Kim.
\newblock Spatial and spectral properties of entangled photons from spontaneous parametric down-conversion with a focused pump.
\newblock {\em Optics Communications}, 366:442--450, 2016.

\bibitem{bennink_optimal_2010}
Ryan~S. Bennink.
\newblock Optimal collinear {Gaussian} beams for spontaneous parametric down-conversion.
\newblock {\em Physical Review A}, 81(5):053805, May 2010.

\bibitem{bernecker_spatial_2023}
Richard Bernecker, Baghdasar Baghdasaryan, and Stephan Fritzsche.
\newblock Spatial and temporal characteristics of spontaneous parametric down-conversion with varying focal planes of interacting beams.
\newblock {\em The European Physical Journal D}, 77(9):172, September 2023.

\bibitem{sevilla-gutierrez_spectral_2024}
Carlos Sevilla-Gutiérrez, Varun~Raj Kaipalath, Baghdasar Baghdasaryan, Markus Gräfe, Stephan Fritzsche, and Fabian Steinlechner.
\newblock Spectral properties of transverse {Laguerre}-{Gauss} modes in parametric down-conversion.
\newblock {\em Physical Review A}, 109(2):023534, February 2024.

\bibitem{grice_spectral_1997}
W.~P. Grice and I.~A. Walmsley.
\newblock Spectral information and distinguishability in type-{II} down-conversion with a broadband pump.
\newblock {\em Physical Review A}, 56(2):1627--1634, August 1997.

\bibitem{ramirez-alarcon_effects_2013}
R.~Ramírez-Alarcón, H.~Cruz-Ramírez, and A.~B. U’Ren.
\newblock Effects of crystal length on the angular spectrum of spontaneous parametric downconversion photon pairs.
\newblock {\em Laser Physics}, 23(5):055204, April 2013.
\newblock Publisher: IOP Publishing.

\bibitem{van_exter_effect_2006}
M.~P. Van~Exter, A.~Aiello, S.~S.~R. Oemrawsingh, G.~Nienhuis, and J.~P. Woerdman.
\newblock Effect of spatial filtering on the {Schmidt} decomposition of entangled photons.
\newblock {\em Physical Review A}, 74(1):012309, July 2006.

\bibitem{rockovich_maximizing_2024}
Andrew Rockovich, Shu'an Wang, and Daniel Gauthier.
\newblock Maximizing {Purity} and {Heralding} {Efficiency} of {Type}-{I} {Down}-{Converted} {Photons} {Using} {Beam} {Focal} {Parameters}, October 2024.
\newblock arXiv:2401.02319 [quant-ph].

\bibitem{zhang_spontaneous_2021}
Chao Zhang, Yun-Feng Huang, Bi-Heng Liu, Chuan-Feng Li, and Guang-Can Guo.
\newblock Spontaneous {Parametric} {Down}-{Conversion} {Sources} for {Multiphoton} {Experiments}.
\newblock {\em Advanced Quantum Technologies}, 4(5):2000132, 2021.
\newblock \_eprint: https://onlinelibrary.wiley.com/doi/pdf/10.1002/qute.202000132.

\bibitem{thekkadath_gain-induced_2024}
Guillaume Thekkadath, Martin Houde, Duncan England, Philip Bustard, Frédéric Bouchard, Nicolás Quesada, and Ben Sussman.
\newblock Gain-{Induced} {Group} {Delay} in {Spontaneous} {Parametric} {Down}-{Conversion}.
\newblock {\em Physical Review Letters}, 133(20):203601, November 2024.

\bibitem{lasota_optimal_2020}
Mikołaj Lasota and Piotr Kolenderski.
\newblock Optimal photon pairs for quantum communication protocols.
\newblock {\em Scientific Reports}, 10(1):20810, November 2020.

\bibitem{takeoka_full_2015}
Masahiro Takeoka, Rui-Bo Jin, and Masahide Sasaki.
\newblock Full analysis of multi-photon pair effects in spontaneous parametric down conversion based photonic quantum information processing.
\newblock {\em New Journal of Physics}, 17(4):043030, April 2015.

\bibitem{mauerer_how_2009}
Wolfgang Mauerer, Malte Avenhaus, Wolfram Helwig, and Christine Silberhorn.
\newblock How colors influence numbers: {Photon} statistics of parametric down-conversion.
\newblock {\em Physical Review A}, 80(5):053815, November 2009.

\bibitem{unternahrer_coincidence_2016}
Manuel Unternährer, Bänz Bessire, Leonardo Gasparini, David Stoppa, and André Stefanov.
\newblock Coincidence detection of spatially correlated photon pairs with a monolithic time-resolving detector array.
\newblock {\em Optics Express}, 24(25):28829, December 2016.

\bibitem{cohen_measuring_2023}
Lior Cohen, Elisha~S. Matekole, Yehuda Pilnyak, Daniel Istrati, Jonathan~P. Dowling, and Hagai~S. Eisenberg.
\newblock Measuring the {Schmidt} number of parametric down conversion by exploiting photon distribution.
\newblock {\em AVS Quantum Science}, 5(2):025002, June 2023.

\bibitem{meher_dependence_2020}
Nilakantha Meher and Anand~K. Jha.
\newblock Dependence of the photon statistics of down-converted field-modes on the photon statistics of the pump field-mode.
\newblock {\em Journal of the Optical Society of America B}, 37(8):2248, August 2020.

\bibitem{waks_direct_2004}
Edo Waks, Eleni Diamanti, Barry~C. Sanders, Stephen~D. Bartlett, and Yoshihisa Yamamoto.
\newblock Direct {Observation} of {Nonclassical} {Photon} {Statistics} in {Parametric} {Down}-{Conversion}.
\newblock {\em Physical Review Letters}, 92(11):113602, March 2004.

\bibitem{glauber_coherent_1963}
Roy~J. Glauber.
\newblock Coherent and {Incoherent} {States} of the {Radiation} {Field}.
\newblock {\em Physical Review}, 131(6):2766--2788, September 1963.
\newblock Publisher: American Physical Society.

\bibitem{paleari_thermal_2004}
Fabio Paleari, Alessandra Andreoni, Guido Zambra, and Maria Bondani.
\newblock Thermal photon statistics in spontaneous parametric downconversion.
\newblock {\em Optics Express}, 12(13):2816, June 2004.

\bibitem{avenhaus_photon_2008}
M.~Avenhaus, H.~B. Coldenstrodt-Ronge, K.~Laiho, W.~Mauerer, I.~A. Walmsley, and C.~Silberhorn.
\newblock Photon {Number} {Statistics} of {Multimode} {Parametric} {Down}-{Conversion}.
\newblock {\em Physical Review Letters}, 101(5):053601, August 2008.

\bibitem{hockel_direct_2011}
David Höckel, Lars Koch, and Oliver Benson.
\newblock Direct measurement of heralded single-photon statistics from a parametric down-conversion source.
\newblock {\em Physical Review A}, 83(1):013802, January 2011.

\bibitem{tapster_photon_1998}
P.~R. Tapster and J.~G. Rarity.
\newblock Photon statistics of pulsed parametric light.
\newblock {\em Journal of modern optics}, 45(3):595--604, 1998.
\newblock Place: ABINGDON Publisher: Taylor \& Francis Group.

\bibitem{waks_highly_2006}
Edo Waks, Barry~C. Sanders, Eleni Diamanti, and Yoshihisa Yamamoto.
\newblock Highly nonclassical photon statistics in parametric down-conversion.
\newblock {\em Physical Review A}, 73(3):033814, March 2006.

\bibitem{galinis_photon_2012}
J.~Galinis, M.~Karpinski, G.~Tamosauskas, K.~Dobek, and A.~Piskarskas.
\newblock {PHOTON} {CORRELATION} {AND} {STATISTICS} {OF} {SPONTANEOUS} {PARAMETRIC} {DOWN}-{CONVERSION} {PUMPED} {BY} {BLUE} {LED} {IN} {LiIO3} {CRYSTAL}.
\newblock {\em Lithuanian journal of physics}, 52(4):285--294, 2012.
\newblock Place: VILNIUS Publisher: Lithuanian Physical Soc.

\bibitem{hamar_non-classical_2014}
M.~Hamar, V.~Michálek, and A.~Pathak.
\newblock Non-classical {Signature} of {Parametric} {Fluorescence} and its {Application} in {Metrology}.
\newblock {\em Measurement Science Review}, 14(4):227--236, August 2014.

\bibitem{wasilewski_statistics_2008}
Wojciech Wasilewski, Czesław Radzewicz, Robert Frankowski, and Konrad Banaszek.
\newblock Statistics of multiphoton events in spontaneous parametric down-conversion.
\newblock {\em Physical Review A}, 78(3):033831, September 2008.

\bibitem{arahata_wavelength_2021}
Masaya Arahata, Yu~Mukai, Bo~Cao, Toshiyuki Tashima, Ryo Okamoto, and Shigeki Takeuchi.
\newblock Wavelength variable generation and detection of photon pairs in visible and mid-infrared regions via spontaneous parametric downconversion.
\newblock {\em Journal of the Optical Society of America B}, 38(6):1934, June 2021.

\bibitem{dorfman_photon_2012}
Konstantin~E. Dorfman and Shaul Mukamel.
\newblock Photon coincidence counting in parametric down-conversion: {Interference} of field-matter quantum pathways.
\newblock {\em Physical Review A}, 86(2):023805, August 2012.

\bibitem{schneeloch_introduction_2019}
James Schneeloch, Samuel~H Knarr, Daniela~F Bogorin, Mackenzie~L Levangie, Christopher~C Tison, Rebecca Frank, Gregory~A Howland, Michael~L Fanto, and Paul~M Alsing.
\newblock Introduction to the absolute brightness and number statistics in spontaneous parametric down-conversion.
\newblock {\em Journal of Optics}, 21(4):043501, April 2019.

\bibitem{osadko_sub-_2007}
I.~S. Osad’ko.
\newblock Sub- and super-{Poissonian} photon statistics of single-molecule fluorescence blinking.
\newblock {\em Journal of Experimental and Theoretical Physics}, 104(6):853--865, June 2007.

\bibitem{trost_photon_2020}
Fabian Trost, Kartik Ayyer, and Henry~N Chapman.
\newblock Photon statistics and signal to noise ratio for incoherent diffraction imaging.
\newblock {\em New Journal of Physics}, 22(8):083070, August 2020.
\newblock Publisher: IOP Publishing.

\bibitem{chan_role_2012}
Kam Wai~Clifford Chan.
\newblock Role of photon statistics of light source in ghost imaging.
\newblock {\em Optics Letters}, 37(13):2739--2741, July 2012.
\newblock Publisher: Optica Publishing Group.

\bibitem{kim_photon-counting_2022}
Jin-Woo Kim, Jeong-Sik Cho, Christian Sacarelo, Nur Duwi~Fat Fitri, Ju-Seong Hwang, and June-Koo~Kevin Rhee.
\newblock Photon-counting statistics-based support vector machine with multi-mode photon illumination for quantum imaging.
\newblock {\em Scientific Reports}, 12(1):16594, October 2022.
\newblock Publisher: Nature Publishing Group.

\bibitem{kuhn_photon-statistics-based_2016}
Simone Kuhn, Sébastien Hartmann, and Wolfgang Elsäßer.
\newblock Photon-statistics-based classical ghost imaging with one single detector.
\newblock {\em Optics Letters}, 41(12):2863--2866, June 2016.
\newblock Publisher: Optica Publishing Group.

\bibitem{ou_multi-photon_2007}
Z.-Y.~J. Ou.
\newblock {\em Multi-photon {Quantum} interference}.
\newblock Springer, 2007.

\bibitem{zmuidzinas_thermal_2003}
J~Zmuidzinas.
\newblock Thermal noise and correlations in photon detection.
\newblock {\em Applied Optics}, 42(25):4989--5008, 2003.
\newblock Place: WASHINGTON Publisher: Optical Soc Amer.

\bibitem{brown_correlation_1956}
R.~Hanbury Brown and R.~Q. Twiss.
\newblock Correlation between {Photons} in two {Coherent} {Beams} of {Light}.
\newblock {\em Nature}, 177(4497):27--29, January 1956.
\newblock Publisher: Nature Publishing Group.

\bibitem{baek_spectral_2008}
So-Young Baek and Yoon-Ho Kim.
\newblock Spectral properties of entangled photon pairs generated via frequency-degenerate type-{I} spontaneous parametric down-conversion.
\newblock {\em Physical Review A}, 77(4):043807, April 2008.

\bibitem{spasibko_spectral_2020}
Kirill Spasibko.
\newblock Spectral and statistical properties of high-gain parametric down-conversion, July 2020.
\newblock arXiv:2007.12999 [quant-ph].

\bibitem{blauensteiner_photon_2009}
Bibiane Blauensteiner, Isabelle Herbauts, Stefano Bettelli, Andreas Poppe, and Hannes Hübel.
\newblock Photon bunching in parametric down-conversion with continuous-wave excitation.
\newblock {\em Physical Review A}, 79(6):063846, June 2009.

\bibitem{fox_quantum_2006}
M.~Fox.
\newblock {\em Quantum {Optics}: {An} {Introduction}}, volume (1st ed., Vol. 15).
\newblock Oxford University Press, 2006.

\end{thebibliography}

\end{document}